\pdfoutput=1

\documentclass[galaxies,article,referee,oneauthor,pdftex,12pt,a4paper]{mdpi}
\usepackage{graphicx}
\makeatletter
\renewcommand\@biblabel[1]{#1. }
\makeatother
\def\h0units{\mathrm{km\,s^{-1}\,Mpc^{-1}}}
\def\cunits{\mathrm{km\,s^{-1}}}
\def\lunits{\mathrm{erg\,s^{-1}}}
\newcommand{\om}{\Omega_{\rm M}}
\newcommand{\ok}{\Omega_K}
\newcommand{\ola}{\Omega_{\Lambda}}

\def\apjl{ApJ\,  }
\def\apjs{ApJS  }

\def\mnras{MNRAS\,  }

\lastpage{x}
\doinum{10.3390/------}
\pubvolume{xx}
\pubyear{2015}
\history{Received: xx / Accepted: xx / Published: xx}
\Title
{
The truncated lognormal distribution as a
luminosity function for  SWIFT-BAT gamma-ray bursts
}

\Author{Lorenzo Zaninetti $^{1,}$}

\address
{
$^{1}$
Physics Department,
 via P.Giuria 1,\\ I-10125 Turin,Italy 
}

\corres
{
zaninetti@ph.unito.it
}

\abstract
{
The determination of the luminosity function (LF)
in gamma ray bursts (GRBs)
depends on the adopted cosmology,
each one characterized by  its corresponding luminosity
distance.
Here we analyse  three cosmologies: the standard cosmology,
the plasma cosmology, and the pseudo-Euclidean universe.
The LF of the GRBs is firstly modeled
by the lognormal distribution 
and the four broken power law,
and secondly
by a truncated lognormal distribution.
The truncated lognormal distribution
fits acceptably the range
in luminosity of GRBs as a function of the redshift.
}



\begin{document}


\section{Introduction}

The number of gamma ray bursts (GRBs)
for which we know the redshift and the flux is 760, according to the
SWIFT-BAT catalog
of \cite{Baumgartner2013}, available at 
the Centre de Donn{\'e}es Astronomiques de Strasbourg (CDS),  
with the name
J/ApJS/207/19.
The above catalog gives the hard X-ray flux,
the spectral index,
the redshift, and the
X-ray luminosity.
The luminosity data of this catalog,
which is a theoretical evaluation, is given 
in the framework of the $\Lambda$CDM cosmology 
with 
$H_0= 70 \h0units$,
$\om=0.3$ and
$\ola=0.7$.
A calibration and a comparison can be done with the
models for luminosity here implemented.
This large number of observed objects allows applying different cosmologies in order to find the luminosity and the
luminosity function (LF) for GRBs.
At the moment of writing, the standard cosmology is the
$\Lambda$CDM cosmology, but other cosmologies
such as the plasma  or the pseudo-Euclidean cosmology
can also be analysed.
Once the luminosity is obtained, we can
model the LF by adopting the lognormal distribution,
see \cite{McBreen1994,Ioka2002}
and by a four broken power law.

In the hypothesis that the luminosity of a GRB is due
to the early phase of a supernova (SN), the minimum 
and maximum are due to the various parameters
which drive the  SN's light curve, 
see \cite{Zaninetti2015c}.

\section{Preliminaries}

\begin{figure}
\begin{center}
\includegraphics[width=10cm]{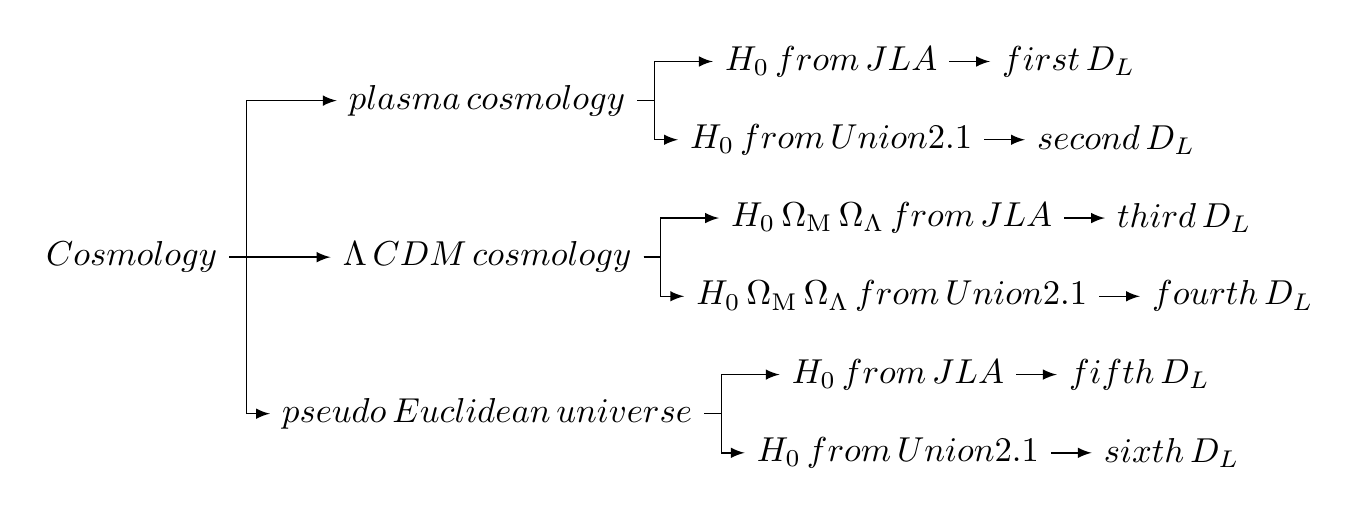}
\end{center}
\caption
{
Flowchart   for the luminosity distances here analysed.
}
\label{forest}
\end{figure}

This section analyses the luminosity in
the $\Lambda$CDM cosmology, in the plasma cosmology
and in the pseudo-Euclidean cosmology.
Careful attention should be paid
to the multiplicative effects
of the main models (3) for the empirical catalogs of SNs (2),
which means 6  different  cases to be analysed,
see Figure  \ref{forest}.

\subsection{Observed luminosity}

In the framework of the standard cosmology,
the  received flux, $f$, is
\begin{equation}
f=\frac{L}{4\,\pi D_L(z)^2}
\quad ,
\end{equation}
where $D_L(z)$ is the luminosity distance, which depends on the
parameters of the adopted cosmological model
and $z$ is the redshift.
As a consequence, the luminosity is
\begin{equation}
L=4\pi
D_L(z)^2 \, f
\quad .
\label{dlzenzakz}
\end{equation}
The above formula  is then corrected by a
$k$-correction, $k(z,\gamma)$,
where
\begin{equation}
k(z,\gamma)= \frac{\int_{1\rm keV}^{10^{4}\rm
keV}C \, E'^{-\gamma} E'dE'}{\int_{15(1+z)\rm keV}^{150(1+z)\rm keV} C \, E'^{-\gamma} E'dE'}
\quad ,
\end{equation}
where $C$ is  a constant and
$\gamma$  is  the observed  spectral index in energy,
see \cite{Tan2013} for more details.
The corrected luminosity is therefore
\begin{equation}
L=4\pi
D_L(z)^2\,  f\, k(z,\gamma)
\quad .
\label{dlkz}
\end{equation}
In the case of the
survey from the 70 month SWIFT-BAT,
the flux $f$ is given in $\frac{fW}{m^2}$,
$\gamma$ and $z$ are  positive numbers,
see \cite{Baumgartner2013};
Table \ref{testgrb}
reports a test-GRB.
\begin{table}[ht!]
\caption
{
Test GRB
}
\label{testgrb}
\begin{center}
\begin{tabular}{|c|c|c|c|c|}
\hline
SWIFT~name   & flux~in~$\frac{fW}{m^2}$  & $\gamma$     & z &
$\log(L(\lunits$))    \\
\hline
J0017.1+8134 &	10.12 &	2.53 &	3.3660&	48.01	 \\
\hline
\end{tabular}
\end{center}
\end{table}

\subsection{Luminosity in the standard cosmology}

The luminosity distance, $D_L$, in the $\Lambda$CDM cosmology can
be expressed in terms of a Pad\'e approximant,
once we provide
the Hubble constant, $H_0$, expressed in  $\h0units$,
the velocity of light, $c$,  expressed in $\cunits$, and
the three numbers $\om$, $\ok$, and $\ola$,
see \cite{Zaninetti2016a} for more details
or Table \ref{lcdmvalues}.
\begin{table}[ht!]
\caption
{
Numerical values of  the $\Lambda$CDM cosmology.
}
\label{lcdmvalues}
\begin{center}
\begin{tabular}{|c|c|c|c|}
\hline
compilation   &  $H_0$ in $\h0units$ & $\om$    & $\ola$    \\
\hline
Union~2.1 & 69.81 & 0.239 & 0.651        \\
\hline
JLA       &  69.398 & 0.181 & 0.538  \\
\hline
\end{tabular}
\end{center}
\end{table}

A further application of the minimax rational approximation,
which is characterized by the two parameters $p$ and $q$,
allows finding a simplified expression
for the luminosity distance,
see eqns~(33a) and (33b) in
\cite{Zaninetti2016a}.
The above minimax approximation
when $p=3,q=2$ is
\begin{equation}
D_{\rm L,3,2}=
\frac
{
p_0  +p_1 \,z +p_2 \,{z}^{2} +p_3\,{z}^{3}
}
{
q_0 +q_1\,z+ q_2 \,{z}^{2}
}\,Mpc
\label{dlzminimax}
\quad ,
\end{equation}
and Table \ref{coefficientspq}
reports the coefficients for the
two compilations here used.

\begin{table}[ht!]
\caption
{
Numerical values of the 7 coefficients
of the minimax approximation
for the Union 2.1 compilation and  the JLA  compilation.
}
\label{coefficientspq}
\begin{center}
\begin{tabular}{|c|c|c|}
\hline
Coefficient  &  Union \, 2.1 & JLA \\
\hline
$p_0$ & 0.3597252600    & 0.4429883062  \\
$p_1$ & 5.612031882     & 6.355991909   \\
$p_2$ & 5.627811123     & 5.405310650   \\
$p_3$ & 0.05479466285   & 0.04413321265 \\
$q_0$ & 0.010587821     & 0.0129850304  \\
$q_1$ & 0.1375418627    & 0.1546989174   \\
$q_2$ & 0.1159043801    & 0.1097492834   \\
\hline
\end{tabular}
\end{center}
\end{table}
The monochromatic luminosity, X-band (14--195 keV),
without $k-z$  correction,
$\log (L_{3,2})_b$
according to eqn~(\ref{dlzenzakz}) is
\begin{multline}
\log (L_{3,2} (\lunits))_b=
\\
 0.43429\,\ln  \left(  1.1964\,{\frac {{\it fluxfwm2}\,
 \left(  16.6843+ \left(  194.6669+ \left(  1878.8341+ 180.34010
\,z \right) z \right) z \right) ^{2}}{ \left(  0.08644+ \left(
 0.2578- 0.00849\,z \right) z \right) ^{2}}} \right)
\\ +
 38.0 \quad Union~2.1
\quad .
\label{lumlcdmmono}
\end{multline}
In the case of a test-GRB with parameters as
in Table \ref{testgrb}, the above formula
gives   $\log (L) = 48.13$  against
$\log (L_{SWIFT}) = 48.01$ of the SWIFT-BAT catalog.
The    goodness of the approximation is evaluated
through the percentage error, $\eta$, which is
\begin{equation}
\eta = \frac{\big | \log (L_{3,2} (\lunits))_b -
\log (L_{SWIFT})\big |}
{ \log (L_{SWIFT)} } \times 100
\quad ,
\end{equation}
and over  all the elements of the SWIFT-BAT catalog
$2.28\,10^{-5} \, \% \leq \eta \leq 0.295 \%$.
We now report an expression for the luminosity of a GRB,
eqn~(\ref{dlkz}), based on the  minimax approximation
when the  Union 2.1 compilation is considered
\begin{multline}
\log (L_{3,2} (\lunits))=
 41.5647+ \\
0.4342\,\ln  \left(  32522\,{\frac {{\it
fluxfwm2}\, \left( z+ 10.3144 \right) ^{2} \left( {z}^{2}+
 0.10378\,z+ 0.0089695 \right) ^{2}}{
 \left( -    0.08644- 0.2578\,z+ 0.008491\,{z}^{2}
 \right) ^{2}}} \right) +\\
\left( {\frac {- 1
+ 0.5
\,\gamma}{ \left(  1 +z \right) ^{2} \left(  \left(  15+ 15\,z
 \right) ^{- \gamma}- 100 \, \left(  150+ 150\,z \right) ^{-
\,\gamma} \right) }} \right)
   \,  Union~2.1
\, \quad ,
\label{loglumlcdm}
\end{multline}
where fluxfwm2  is the flux expressed
in $\frac{fW}{m^2}$.

\begin{figure}
\begin{center}
\includegraphics[width=10cm]{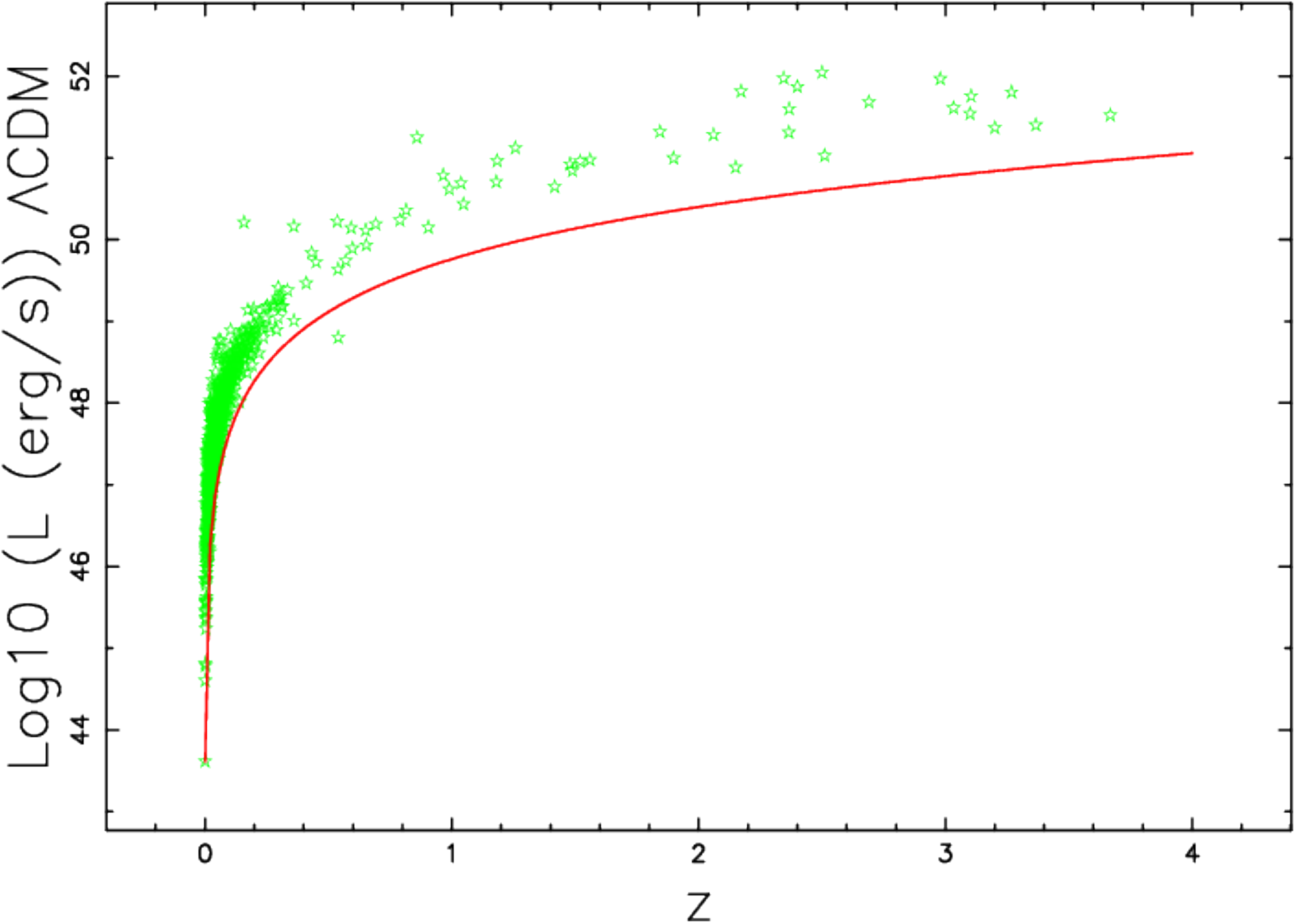}
\end{center}
\caption
{
Luminosity in the $\Lambda$CDM cosmology
versus redshift
for 784 GRB
as given by the 70 month SWIFT-BAT survey (green points)
and theoretical curve for the lowest luminosity at  a given
redshift (red curve),
see eqn~(\ref{loglumlcdm}).
}
\label{lumin_lcdm}
\end{figure}

In the case of a test-GRB with parameters as
in Table \ref{testgrb}, the above formula
gives  $\log (L) = 54.512$
which means  a bigger luminosity of $\approx$ 6 decades
with respect to the band luminosity.
Figure \ref{lumin_lcdm}
reports the
luminosity--redshift
distribution for the SWIFT-BAT survey as well a
a theoretical lower curve which can be found
by inserting the minimum flux in
eqn~(\ref{loglumlcdm}).
Another useful quantity is the angular 
diameter distance, $D_A$,
which is 
\begin{equation}
D_A = \frac{D_L}{(1+z)^2}
\quad ,
\end{equation}
see \cite{Etherington1933}, and therefore
\begin{equation}
D_{\rm A,3,2} = \frac{ D_{\rm L,3,2}}{(1+z)^2}
\quad .
\label{dazminimax}
\end{equation}

\subsection{Luminosity in the plasma cosmology}

The distance  $d$ in the plasma cosmology
has   the  following dependence:
\begin{equation}
d(z)=
{\frac {\ln  \left( z+1 \right) c}{H_{{0}}}}
\quad ,
\label{distancenltired}
\end{equation}
 see
\cite{Brynjolfsson2004,Ashmore2006,Zaninetti2015a,Ashmore2016}
and Table \ref{h0plasma}. 
\begin{table}[ht!]
\caption
{
Numerical values of  $H_0$  in $\h0units$ (plasma cosmology)
for the Union 2.1 compilation and  the JLA  compilation.
}
\label{h0plasma}
\begin{center}
\begin{tabular}{|c|c|}
\hline
Union \, 2.1 & JLA \\
$H_0=74.2\pm 0.24 $ & $H_0=74.45\pm 0.2$ \\
\hline
\end{tabular}
\end{center}
\end{table}
The monochromatic luminosity, X-band (14--195 keV), is
\begin{equation}
\log(L(z))={\frac {\ln  \left(  19531902.82\,{\it fluxfwm2}\, \left( \ln  \left(
1+z \right)  \right) ^{2} \right) }{\ln  \left( 10 \right) }}+38
\quad .
\label{loglumplasmasenza}
\end{equation}
In the case of a test-GRB with parameters as
in Table \ref{testgrb}, the above formula
gives   $\log (L) = 46.63$, which a lower value than the
$\log (L_{SWIFT}) = 48.01$  of the SWIFT-BAT catalog.

\begin{figure}
\begin{center}
\includegraphics[width=10cm]{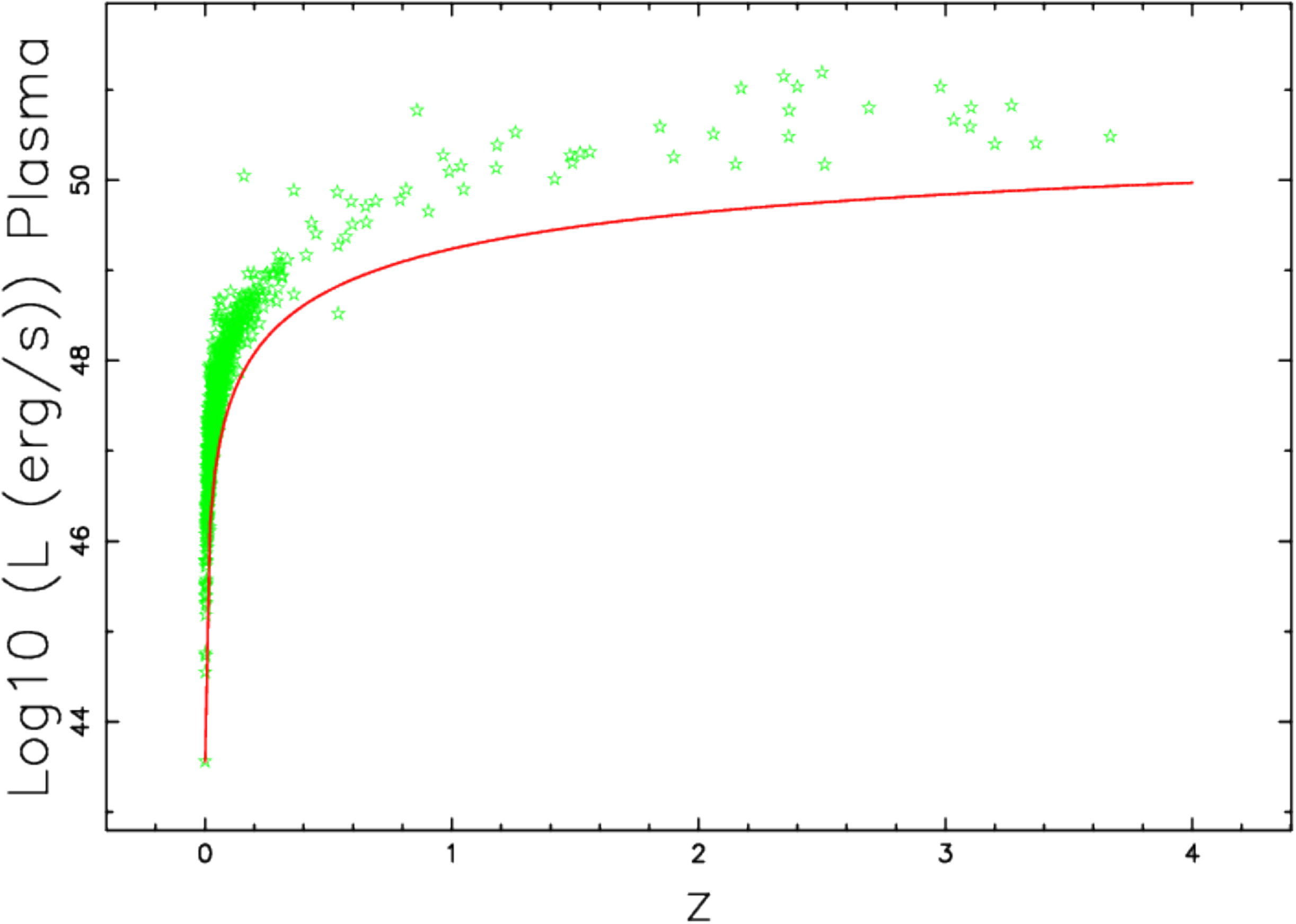}
\end{center}
\caption
{
Luminosity in the plasma  cosmology
versus redshift
for 784 GRB
as given by the 70 month SWIFT-BAT survey (green points)
and theoretical curve for the lowest luminosity at  a given
redshift (red curve),
see eqn~(\ref{luminosityplasma}).
}
\label{lumin_plasma}
\end{figure}

The luminosity in the case of the absence of absorption  is
\begin{equation}
L(z)=4\pi
d(z)^2 \, f \, k(\gamma)
\label{lumplasmakz}
\quad ,
\end{equation}
where the $k(\gamma)$ correction is
\begin{equation}
k(\gamma)= \frac{\int_{1\rm keV}^{10^{4}\rm
keV}C \, E'^{-\gamma} E'dE'}{\int_{15\rm keV}^{150\rm keV} C \, E'^{-\gamma} E'dE'}
\quad .
\label{luminosityplasma}
\end{equation}
There is no relativistic correction in the denominator because
the plasma cosmology is both static and Euclidean.
Figure \ref{lumin_plasma}
reports the luminosity in the plasma cosmology
as a function of the redshift
as well
as the theoretical luminosity.

\subsection{Luminosity in the pseudo-Euclidean cosmology}

The distance  $d$  in the pseudo-Euclidean cosmology
has   the  following dependence:
\begin{equation}
d(z)=
\frac {z c}{H_{{0}}}
\quad ,
\label{distancepseudo}
\end{equation}
and we used $H_0=67.93 \h0units $, see Table
\ref{h0pseudoeu}.
 \begin{table}[ht!]
\caption
{
Numerical values of  $H_0$  in $\h0units$
(pseudo-Euclidean cosmology)
for the Union 2.1 compilation and  the JLA  compilation when the
redshift covers the range $[ 0, 0.1]$
}
\label{h0pseudoeu}
\begin{center}
\begin{tabular}{|c|c|}
\hline
Union \, 2.1 & JLA \\
$H_0=67.93 \pm 0.38 $ & $H_0= 67.51\pm 0.42$ \\
\hline
\end{tabular}
\end{center}
\end{table}

The above formula gives  approximate results up
to $z \ll 1.0 $.
The monochromatic luminosity, X-band (14--195 keV),  is
\begin{equation}
L(z)=4\pi
d(z)^2 \, f
\quad ,
\label{lumpseudosenza}
\end{equation}
where the $k(z)$ correction is absent or
\begin{equation}
\log(L(z))= {\frac {\ln  \left(  19531902.82\,{\it fluxfwm2}\,{z}^{2} \right) }{
\ln  \left( 10 \right) }}+38
\quad .
\end{equation}

\subsection{High versus low $z$}

\begin{figure}
\begin{center}
\includegraphics[width=10cm]{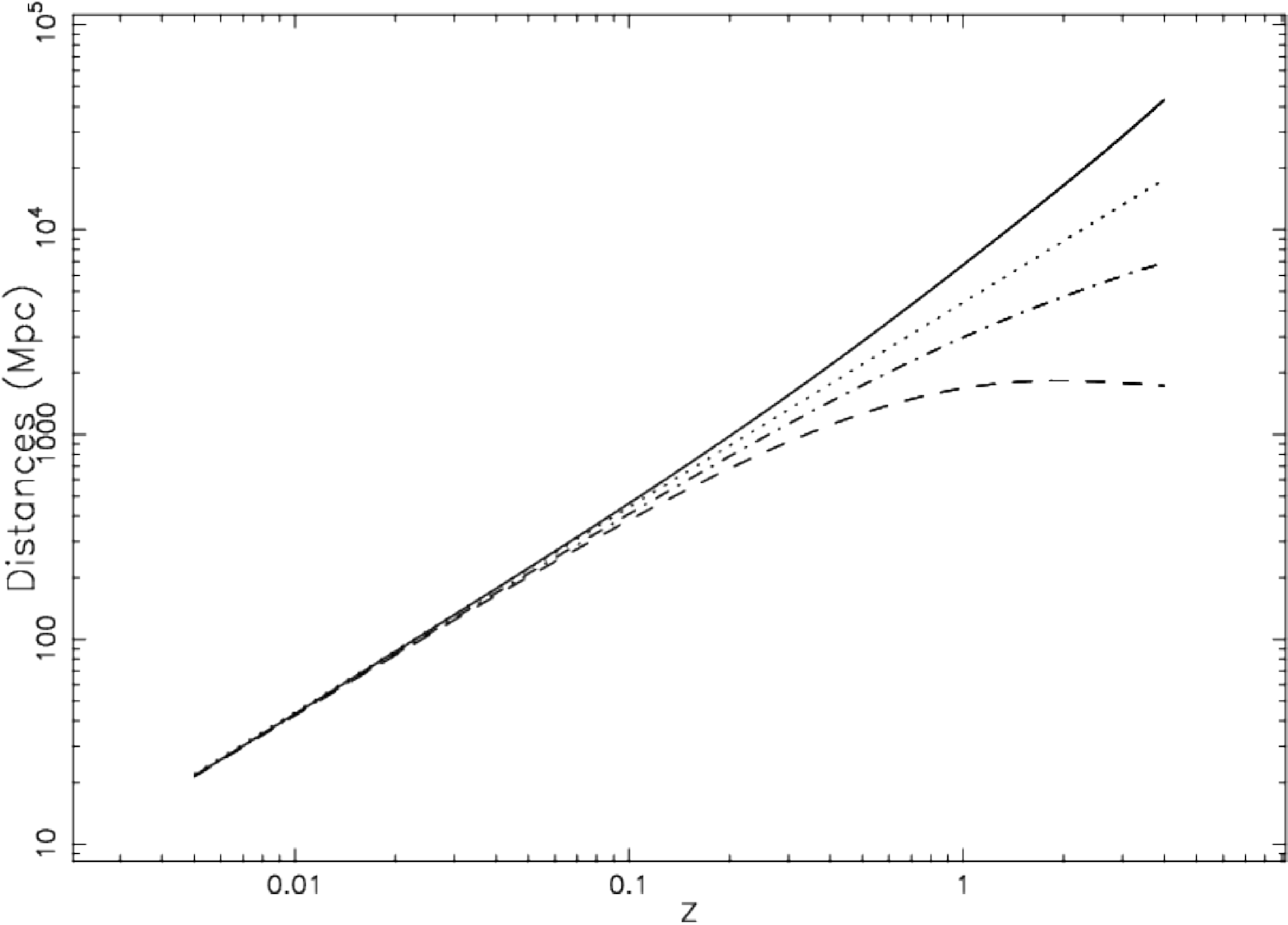}
\end{center}
\caption
{
The  distances here adopted:
luminosity distance,        $D_L$, in $\Lambda$CDM (full line),
angular-diameter  distance, $D_A$, in $\Lambda$CDM (dash line),
plasma cosmology  distance, $d$, (dot-dash-dot-dash line) 
and 
pseudo-Euclidean cosmology distance (dotted line). 
}
\label{distances_z}
\end{figure}

The differences between the four distances here used,
which are the 
luminosity distance and the
angular-diameter  distance in the $\Lambda$CDM,
the plasma cosmology  distance, 
and the 
pseudo-Euclidean cosmology distance, can be outlined
in terms of a percentage difference,
 $\Delta$.
As an example for $D_{\rm A}$,
\begin{equation}
\Delta = \frac{\big | D_{\rm L}(z) - D_{\rm A}(z) \big |}
{D_{\rm L}(z)} \times 100
\quad .
\end{equation}
Figure \ref{distances_z} reports the four distances 
and for $z\leq 0.05$ the three percentage differences are 
lower than $10\%$.
In the framework of the two Euclidean distances, the plasma 
and the pseudo-Euclidean one,
for $z\leq 0.15$ the percentage difference is 
lower than $10\%$. 
Therefore the boundary between low and high $z$ can be fixed 
at $z=0.05$.

\section{Two existing  distributions}

This section reviews the four broken power law distribution
and the lognormal distribution
and derives an analytical expression for
the number of GRBs for a given flux in
the linear and non-linear cases.

\subsection{The four broken power law distribution}

The four broken power law has  the  following piecewise dependence:
\begin{equation}
p(L)  \propto L^{\alpha_i} \quad,
\end {equation}
each of the four  zones  being  characterized  by a different
exponent  ${\alpha_i}$.
In order to have a PDF normalized  to unity, one must have
\begin{equation}
\sum _{i=1,4}  \int_{L_i}^{L_{i+1}} c_i L ^{\alpha_i} dL =1
\quad.
\label{uno}
\end{equation}
For example,  we start with $c_1$=1:
$c_2$   will be determined by
the following
equation:
\begin{equation}
c_1 (L_2 - \epsilon)^{\alpha_1} =
c_2 (L_2 + \epsilon)^{\alpha_2}
\quad,
\end{equation}
where $\epsilon$ is a small number, e.g.
$\epsilon =\frac{L_2}{10^{+8}}$.
This  PDF is characterized by 9  parameters 
and takes 
values $L$ in the interval
$[L_1, L_5]$.

\subsection{Lognormal distribution}
\label{seclognormal}

Let $L$ be a random variable taking
values $L$ in the interval
$[0, \infty]$; the {\em lognormal}
probability density
function (PDF),
following \cite{evans}
or formula (14.2)$^\prime$ in
\cite{univariate1}, is
\begin{equation}
PDF (L;L^*,\sigma) = \frac
{
\sqrt {2}{{\rm e}^{-\frac{1}{2}\,{\frac {1}{{\sigma}^{2}} \left( \ln  \left( {
\frac {L}{{\it L^*}}} \right)  \right) ^{2}}}}
}
{
2\,L\sigma\,\sqrt {\pi }
}
\quad,
\label{pdflognormal}
\end{equation}
where
$L^*=\exp{\mu_{LN}}$ and
$\mu_{LN}=\ln {L^*}$.
The mean luminosity is
\begin{equation}
E(L;L^*,\sigma)=
{\it L^*}\,{{\rm e}^{\frac{1}{2}\,{\sigma}^{2}}}
\quad ,
\end{equation}
and the  variance, $Var$, is
\begin{equation}
Var(L^*,\sigma)={{\rm e}^{{\sigma}^{2}}} \left( -1+{{\rm e}^{{\sigma}^{2}}} \right) {{
\it L^*}}^{2}
\quad .
\end{equation}
The distribution function (DF)  is
\begin{equation}
DF(L;L^*,\sigma)=\frac{1}{2}+ \frac{1}{2}\,{\rm erf}
\left(\frac{1}{2} \,{\frac {\sqrt {2} \left( \ln  \left( L
 \right) -\ln  \left( {\it L^*} \right)  \right) }{\sigma}}\right)
\quad ,
\end{equation}
where     ${\rm erf(z)}$ is the error function,
see \cite{NIST2010}.
A luminosity function for GRB, $PDF_{GRB}$,
can be obtained
by  multiplying  the lognormal PDF by
$\Phi^*$, which is the number of GRB per unit volume,
Mpc$^3$ units   for unit time,  yr units,
\begin{equation}
\Phi (L;L^*,\sigma) =\Phi^* \frac
{
\sqrt {2}{{\rm e}^{-\frac{1}{2}\,{\frac {1}{{\sigma}^{2}} \left( \ln  \left( {
\frac {L}{{\it L^*}}} \right)  \right) ^{2}}}}
}
{
2\,L\sigma\,\sqrt {\pi }
}
\frac{number}{Mpc^3 \,yr}
\quad .
\label{pdflognormalgrb}
\end{equation}

\begin{figure}
\begin{center}
\includegraphics[width=10cm]{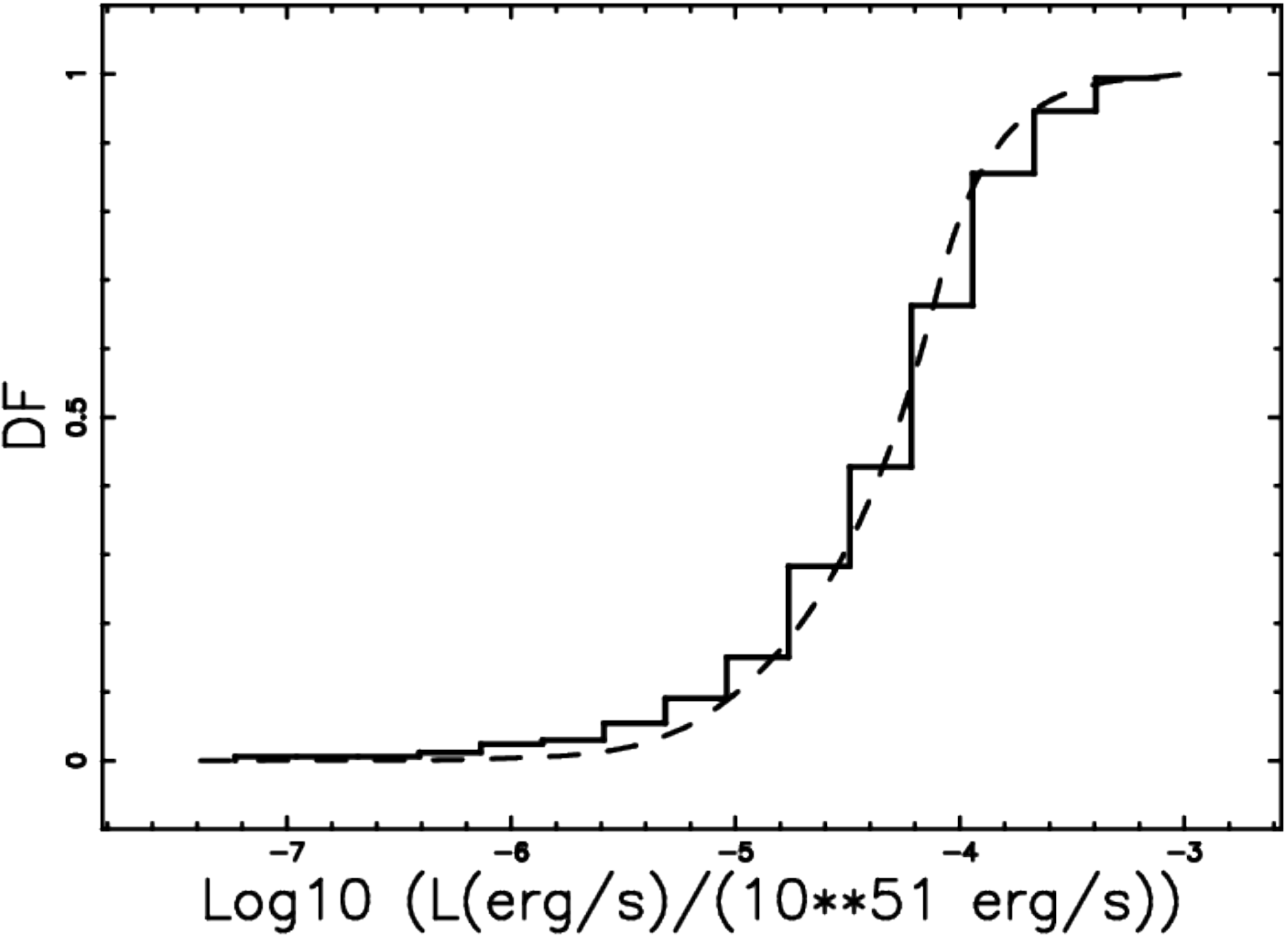}
\end {center}
\caption
{
Observed DF (step-diagram)
for  GRB luminosity and  superposition
of the four broken power laws'  DFs  (line),
case of $\Lambda$CDM cosmology with  parameters as
in Table \ref{fourbroken}.
}
\label{quattro_df}
    \end{figure}

A numerical value for the  constant
$\Phi^*$  can be obtained by dividing
the number of GRB, $N_{GRB}$,  observed in a time, $T$,
in a given volume  $V$ by the volume itself
and by  $T$,  
which is the time over which the
phenomena are observed, in the case  of
SWIFT-BAT,  70 month,
see \cite{Baumgartner2013},
\begin{equation}
\Phi^* = \frac{N_{GRB}}{V\,T} Mpc^{-3} yr^{-1}
\label{fistar}
\quad ,
\end{equation}
where the volume is different in the three cosmologies,
\begin{subequations}
\begin{align}
V  =&  \frac{4}{3} \pi (\frac{c z}{H_0})^3  \, Mpc^3 &
pseudo-Euclidean \,cosmology
\label{vpseudoeuclidean}     \\
V  =&  \frac{4}{3} \pi
({\frac {\ln  \left( z+1 \right) c}{H_{{0}}}} )^3 \, Mpc^3
& plasma \, cosmology \label {vplasma}  \\
V  =&  \frac{4}{3} \pi
(D_{\rm A,3,2})^3 \, Mpc^3
&  \Lambda\,CDM \,cosmology \label {vstandard}
\quad ,
\end{align}
\end{subequations}
where $D_{\rm A,3,2}$ has been defined 
in eqn~(\ref{dazminimax}).
The  parameters of the fit  for  the  four broken power law's 
PDF are reported in
Table \ref{fourbroken} when the luminosity
is taken with the $k(z)$ correction,
Figure \ref{quattro_df}.

\begin{table}[ht!]
\caption 
{
The 9 parameters of the four broken power laws 
for 
the $\Lambda$CDM  cosmology
where 
eqn~(\ref{loglumlcdm}) was used 
and  the two parameters of the K--S test
$D$ and  $P_{KS}$.
}
\label{fourbroken}
\begin{center}
\begin{tabular}{|c|c|}
\hline
Name           & \\
\hline  
$L_1$\,in\,$\frac{L^*}{10^{51} erg\,s^{-1}}$          & $4  \,10^{-8}$                       \\
$L_2$\,in\,$\frac{L^*}{10^{51} erg\,s^{-1}}$          & $5  \,10^{-7}$                       \\
$L_3$\,in\,$\frac{L^*}{10^{51} erg\,s^{-1}}$          & $6.3\,10^{-6}$                       \\
$L_4$\,in\,$\frac{L^*}{10^{51} erg\,s^{-1}}$          & $7.9\,10^{-5}$                       \\
$L_5$\,in\,$\frac{L^*}{10^{51} erg\,s^{-1}}$          & $9.8\,10^{-4}$                       \\
$\alpha_1$     & $1.2$                                \\
$\alpha_2$     & $0.54$                                \\
$\alpha_3$     & $-0.23$                                \\
$\alpha_4$     & $-2.74$                                \\
$D$            & 0.063                                 \\
$ P_{KS}$      & 0.507                                 \\
\hline                                                   
\end{tabular}
\end{center}
\end{table}

The  parameters of the fit  for  the  lognormal
PDF are reported in
Table \ref{coeflognormal} when the luminosity
is taken with the $k(z)$ correction.
\begin{table}[ht!]
\caption
{
The 3 parameters of  the LF as modeled by the
lognormal distribution
for  $z$ in $[0,0.02]$ with the Union~2.1 data
and  the two parameters of the K--S test
$D$ and  $P_{KS}$.
In the case of
the plasma cosmology
and
the $\Lambda$CDM  cosmology
we used
the luminosity as  given by
eqn~(\ref{lumplasmakz})
and
eqn~(\ref{loglumlcdm}), respectively.
}
\label{coeflognormal}
\begin{center}
\begin{tabular}{|c|c|c|c|}
\hline
Parameter  &  Plasma~cosmology & $\Lambda$CDM cosmology \\
\hline
$\frac{L^*}{10^{51} erg\,s^{-1}}$    & 3.516\,$10^{-5}$
& 4.055\,$10^{-5}$   \\
$\sigma$ & 1.42                                 & 1.42   \\
$\frac
{\Phi^*}
{Mpc^{-3} yr^{-1}}
$ & $7.2524\,10^{-8}$                              & $1.025 \,10^{-5}$
\\
$D$       & 0.089     & 0.090   \\
$ P_{KS}$ & 0.131     & 0.127   \\
\hline
\end{tabular}
\end{center}
\end{table}

The case of LF modeled by a  lognormal PDF
with $L$  as represented  by
a  monochromatic luminosity in the X-band (14--195 keV)
is reported in Table \ref{coeflognormalsenza}.

\begin{table}[ht!]
\caption
{
The 3 parameters of  the LF,  case of  X-band (14--195 keV),
as modeled by the lognormal distribution
for  $z$ in $[0,0.02]$ with the Union~2.1 data
and  the two parameters of the K--S test,
$D$ and  $ P_{KS}$.
In the case of the plasma cosmology and the pseudo-Euclidean cosmology,
we used
the luminosity as  given by
eqn~(\ref{loglumplasmasenza})
and
eqn~(\ref{lumpseudosenza}),
respectively.
}
\label{coeflognormalsenza}
\begin{center}
\begin{tabular}{|c|c|c|}
\hline
Parameter  &  Plasma~cosmology &
 pseudo-Euclidean cosmology\\
\hline
$\frac{L^*}{10^{51} erg\,s^{-1}}$    & 5.9\,$10^{-9}$
& 7.12\,$10^{-9}$   \\
$\sigma$ & 1.42                                 & 1.42   \\
$\frac
{\Phi^*}
{Mpc^{-3} yr^{-1}}
$ & $1.01\,10^{-5}$                              & $9.88 \,10^{-6}$
\\
$D$       & 0.089    & 0.089   \\
$ P_{KS}$ & 0.13     & 0.129   \\
\hline
\end{tabular}
\end{center}
\end{table}
The goodness of the fit with the  lognormal  PDF  has been
assessed by  applying the Kolmogorov--Smirnov  (K--S) test
\citep{Kolmogoroff1941,Smirnov1948,Massey1951}.
The K--S test,
as implemented by the FORTRAN subroutine KSONE in \cite{press},
finds
the maximum  distance, $D$, between the theoretical
and the observed  DF
as well the  significance  level,   $P_{KS}$,
see formulas  14.3.5 and 14.3.9  in \cite{press};
values of $ P_{KS} \geq 0.1 $ indicate that the fit is acceptable,
see Table \ref{coeflognormal} for the results.

\begin{figure}
\begin{center}
\includegraphics[width=10cm]{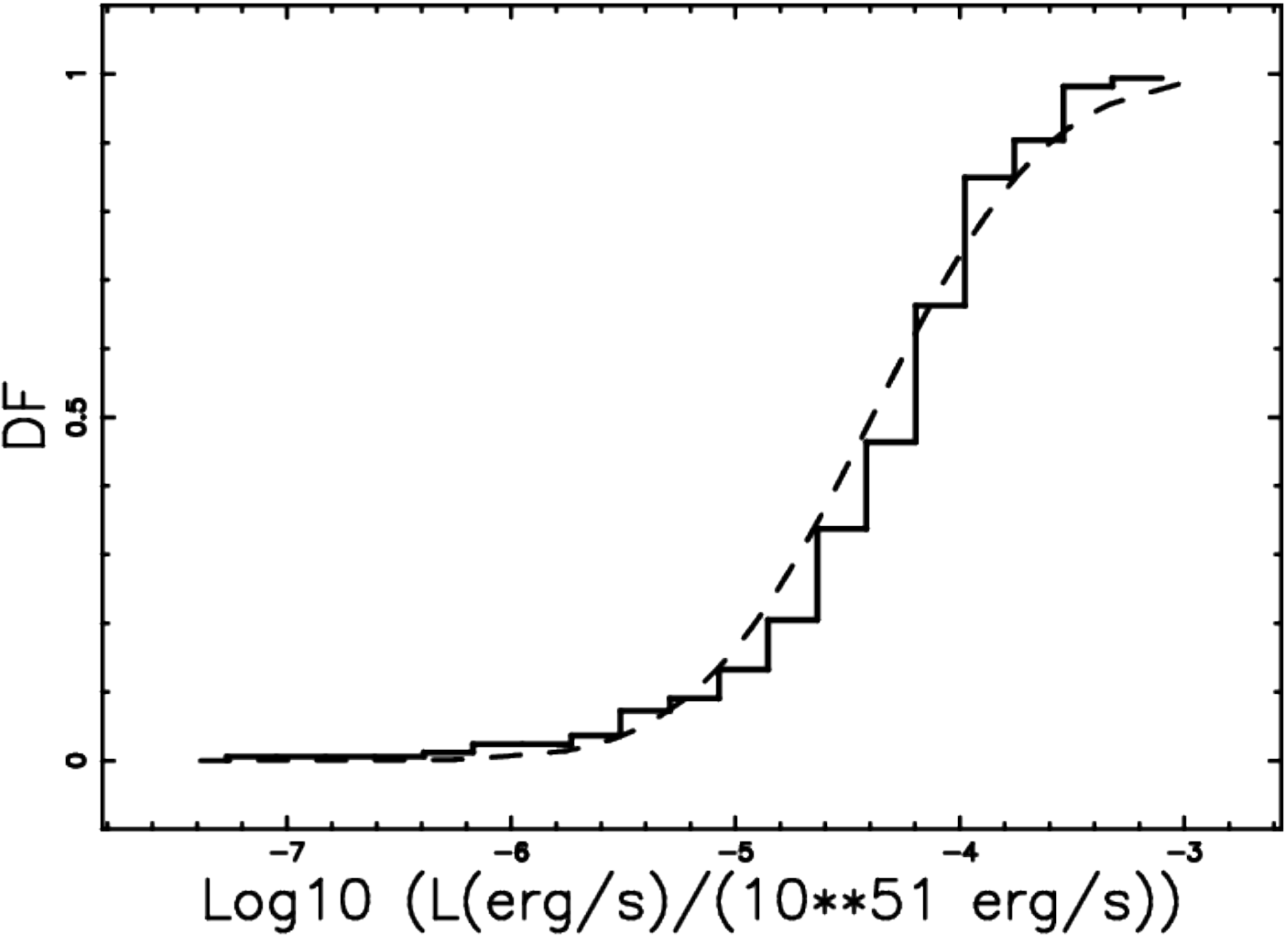}
\end {center}
\caption
{
Observed DF (step-diagram)
for  GRB luminosity and  superposition
of the lognormal DF  (line),
case of the $\Lambda$CDM cosmology with  parameters as
in Table \ref{coeflognormal}.
}
\label{lognormal_df}
    \end{figure}

In the case of the $\Lambda$CDM  cosmology
Figure \ref{lognormal_df}
reports the   lognormal DF,
with parameters as in Table \ref{coeflognormal}.

\begin{figure}
\begin{center}
\includegraphics[width=10cm]{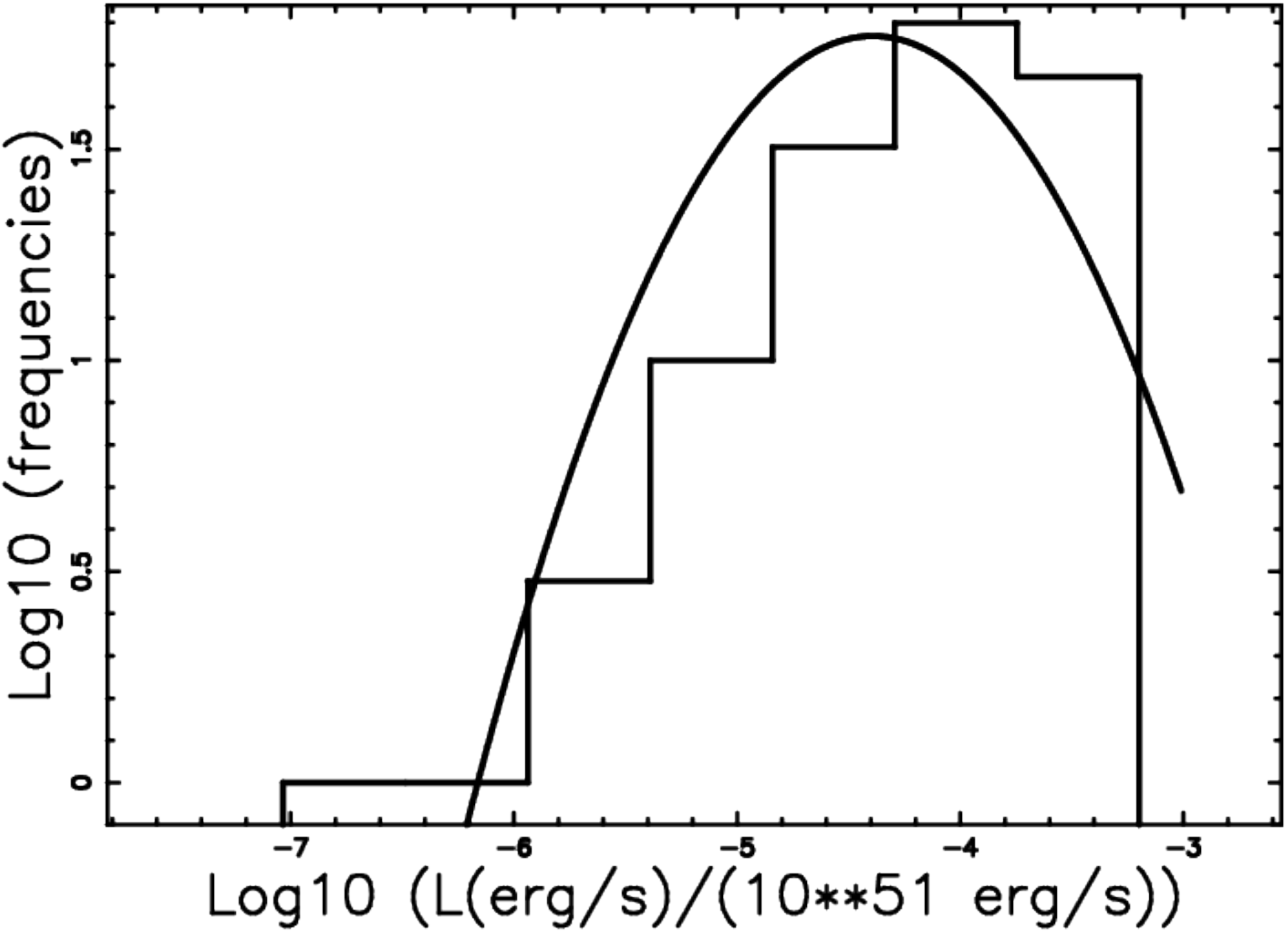}
\end {center}
\caption
{
Log--Log  histogram     (step-diagram)
of GRB luminosity and  superposition
of the lognormal PDF  (line),
case of the pseudo-Euclidean cosmology
with  parameters as
in Table \ref{coeflognormal}.
}
\label{lognormal_isto}
    \end{figure}

In the case of the $\Lambda$CDM  cosmology,
Figure \ref{lognormal_isto}
reports
a comparison between
the empirical distribution and
the   lognormal  PDF,
and Figure \ref{lognormal_df}
reports
the   lognormal DF,
with parameters as in Table \ref{coeflognormal}.

\begin{figure}
\begin{center}
\includegraphics[width=10cm]{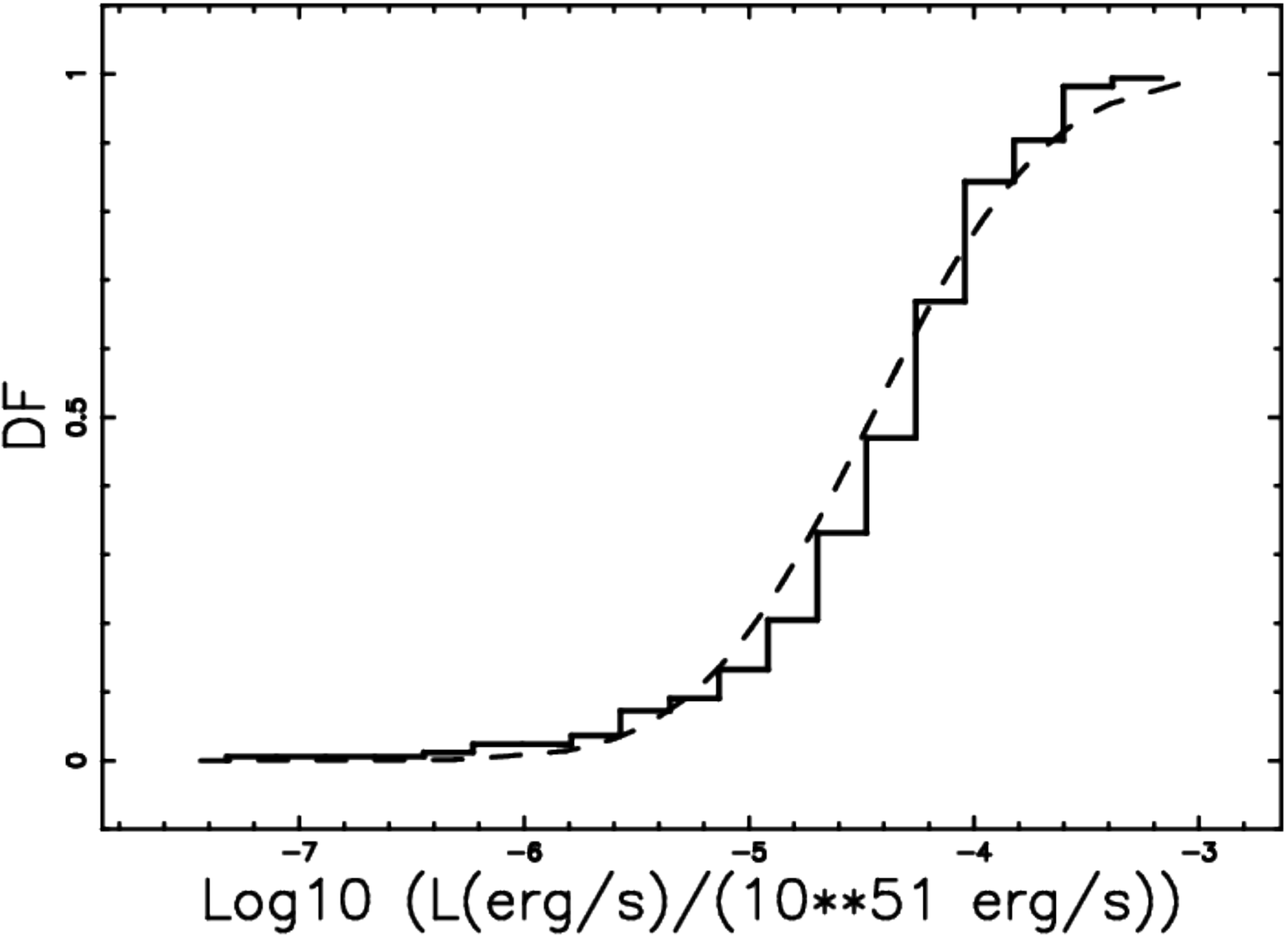}
\end {center}
\caption
{
Observed DF (step-diagram)
for  GRB luminosity and  superposition
of the lognormal DF  (line),
case of the plasma cosmology with  parameters as
in Table \ref{coeflognormal}.
}
\label{lognormal_df_plasma}
    \end{figure}

\begin{figure}
\begin{center}
\includegraphics[width=10cm]{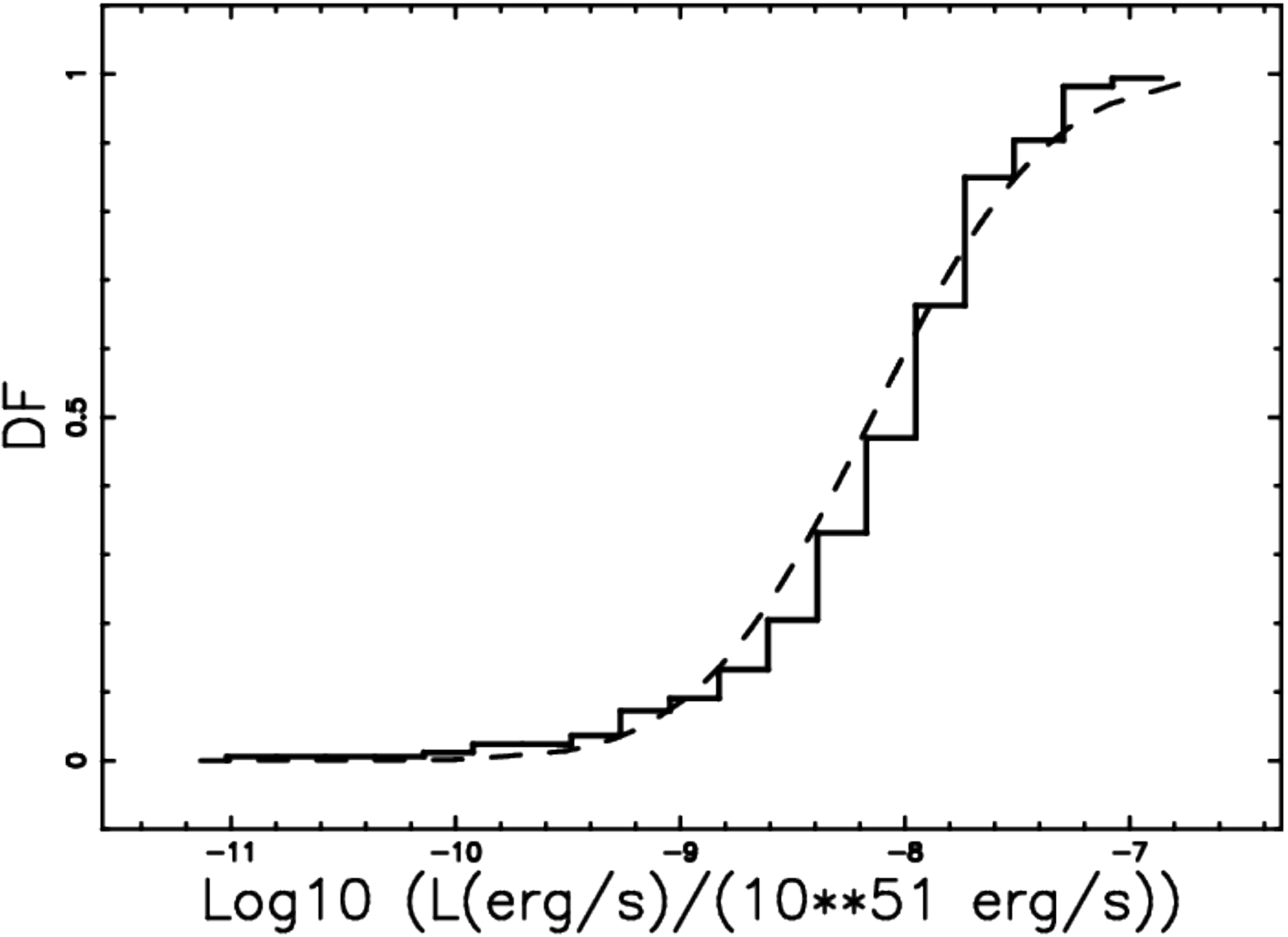}
\end {center}
\caption
{
Observed DF (step-diagram)
for  GRB monochromatic luminosity, X-band (14--195 keV),
and  superposition
of the lognormal DF  (line),
case of the pseudo-Euclidean cosmology
with  parameters as
in Table \ref{coeflognormal}.
}
\label{lognormal_df_pseudoeu}
    \end{figure}

The case of the plasma  and pseudo-Euclidean cosmologies are covered
in Figs 
\ref{lognormal_df_plasma} and 
\ref{lognormal_df_pseudoeu}
respectively.

\subsection{The linear case}

 We assume that the flux, $f$,
 scales as  $f  = \frac{L}{4 \pi r^2}$,
according to eqn~(\ref{distancepseudo}):
\begin{equation}
 r =
\frac {z c}{H_{{0}}}
\quad ,
\end{equation}
and
\begin{equation}
z={\frac {{\it r}\,H_{{0}}}{c}}
\quad .
\end{equation}

\begin{figure}
\begin{center}
\includegraphics[width=6cm]{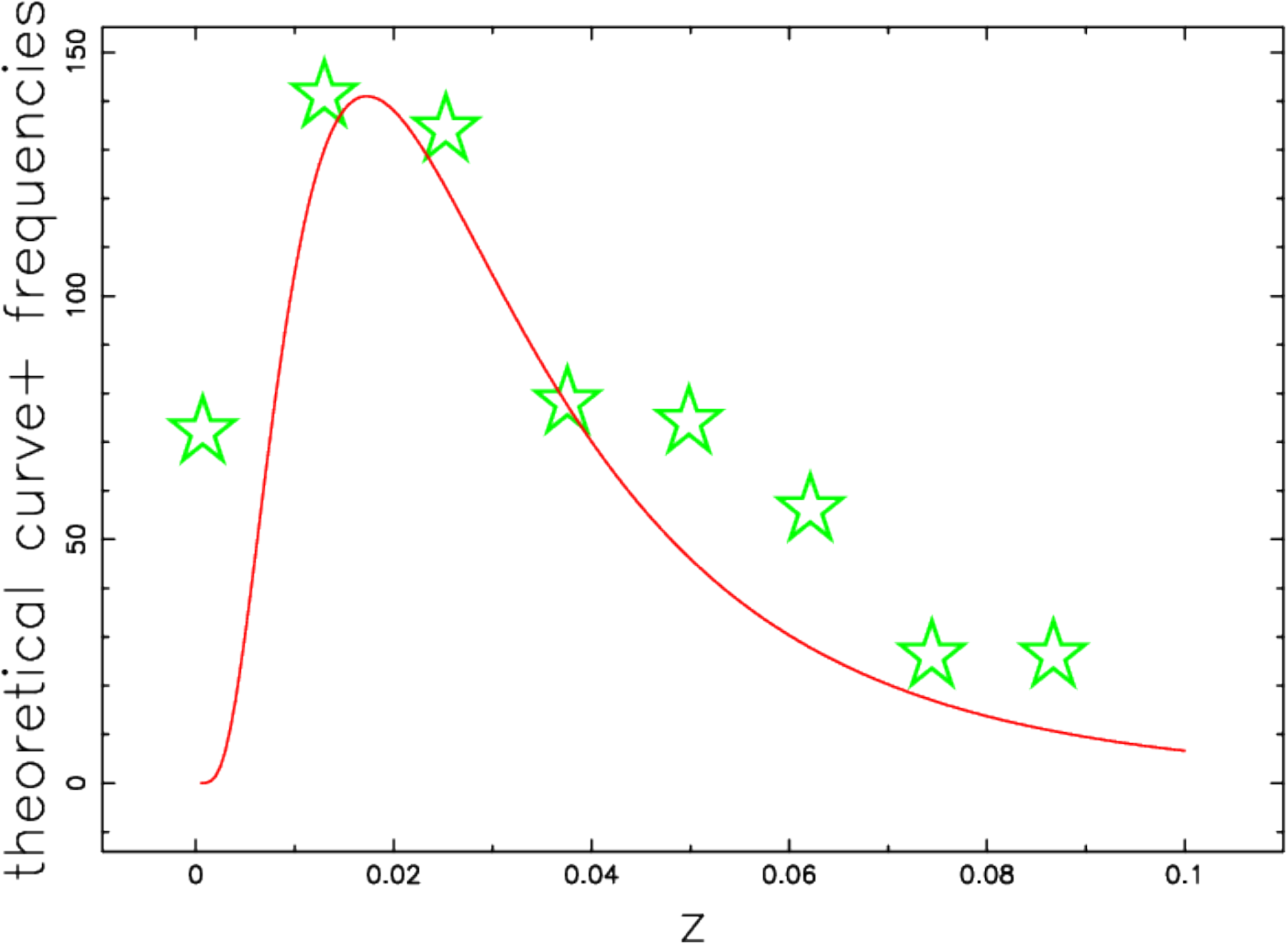}
\end {center}
\caption
{
The GRBs of  the SWIFT-BAT catalog    with
$ 3.1 \,\frac{fW}{m^2}
\leq  f
\leq 150.54\, \frac{fW}{m^2}$,
which means $\langle f \rangle =76.82 \, \frac{fW}{m^2} $,
are  organized in frequencies versus
spectroscopic   redshift (green stars).
The redshift covers the range $[ 0, 0.1]$,
the maximum frequency in the observed GRbs  is
at $z=0.019$, $\chi^2 =  5925$
and the number of bins is 8.
The full red line is the theoretical curve
generated by
$\frac{dN}{d\Omega dz df}(z)$
as given by the application of the lognormal LF
which  is eqn~(\ref{dnsudz_pseudo}) in the
pseudo-Euclidean cosmology with
parameters as in Table \ref{coeflognormal}.
}
          \label{max_pseudoeu}%
    \end{figure}

The relation between the two differentials $dr$
and $dz$ is
\begin{equation}
dr = {\frac {{\it c
}\,{\it dz}}{H_{{0}}}}
\quad .
\end{equation}
The joint distribution in {\it z}
and  {\it f}  for the number of galaxies
 is
\begin{equation}
\frac{dN}{d\Omega dz df} =
\frac{1}{4\pi}\int_0^{\infty} 4 \pi r^2 dr \Phi (\frac{L}{L^*})
\delta\bigl(z- ({\frac {{\it r}\,H_{{0}}}{c}})\bigr)
\delta\bigl(f-\frac{L}{4 \pi r^2}    \bigr)
\quad ,
\label{nfunctionzschechternonldef}
\end{equation}
where $\delta$ is the Dirac delta function.
We now introduce the  critical value of 
$z$,   
$z_{crit}$, which is
\begin{equation}
 z_{crit}^2 = \frac {H_0^2  L^* } {4 \pi f c^2}
\quad .
\end{equation}
The evaluation of the integral  over luminosities and distances
gives
\begin{equation}
\frac{dN}{d\Omega dz df} = F(z;f,\Phi^*,L^*,\sigma) =
\frac
{
{z}^{2}{c}^{3}\sqrt {2}{{\rm e}^{-\frac{1}{2}\,{\frac {1}{{\sigma}^{2}}
 \left( \ln  \left( {\frac {{z}^{2}}{{z_{{{\it crit}}}}^{2}}} \right)
 \right) ^{2}}}}{\it \Phi^*}
}
{
2\,\sqrt {\pi}{H_{{0}}}^{3}f\sigma
}
\label{dnsudz_pseudo}
\quad ,
\end{equation}
where $d\Omega$, $dz$ and  $ df $ represent
the differential of
the solid angle,
the redshift, and the flux, respectively,
and
$\Phi^*$ is the normalization of the lognormal LF for GRB.
The number of GRBs in $z$  and $f$ as given by the above
formula has a maximum  at  $z=z_{pos-max}$,
where
\begin{equation}
 z_{pos-max} = {{\rm e}^{\frac{1}{2}\,{\sigma}^{2}}}z_{{{\it crit}}}
\label{zmaxpseudo}
\quad ,
\end{equation}
which  can be re-expressed   as
\begin{equation}
 z_{pos-max} =
 \frac{{{\rm e}^{\frac{1}{2}\,{\sigma}^{2}}}\sqrt {{\it L^*}}H_{{0}}}{2\,\sqrt {\pi}\sqrt {f}c}
 \quad .
 \end{equation}
Figure \ref{max_pseudoeu}
reports the
observed and theoretical number of GRBs
with a given flux
as a function of the  redshift.
The theoretical maximum
as  given by eqn~(\ref{zmaxpseudo})
is  at  $z=0.017$, with the parameters as in Table \ref{coeflognormal},
against  the observed $z=0.019$.
The theoretical mean redshift of GRBs  with
flux  $f$  can be deduced from
eqn~(\ref{dnsudz_pseudo}):
\begin{equation}
\langle z \rangle =\frac{\int_0^{\infty} z \,F(z;f,L^*,\Phi^*,\sigma) dz}
{\int_0^{\infty} F(z;f,L^*,\Phi^*,\sigma) dz}
\quad .
\label{zaverage}
\end{equation}
The above integral does not have an analytical expression,
and should  be numerically evaluated.
The above formula with parameters as in
Figure \ref{max_pseudoeu} gives a theoretical/numerical
$\langle z \rangle= 0.0368$
 against the observed
$\langle z \rangle= 0.0385$.
The quality of the fit  in the number of
GRBs with a given flux depends on the 
chosen flux, the interval of the flux in which the frequencies 
are evaluated, and the number of histograms.
A larger number of available GRBs will presumably increase
the goodness of the fit.

\subsection{The non-linear case}

We assume that  $f  = \frac{L}{4 \pi r^2}$
and
\begin{equation}
 z =e^{(H_0 \, r/c )} -1
\quad ,
\end{equation}
where $r$ is the distance;
 in our case, $d$ is as represented
 by the non-linear  eqn~(\ref{distancenltired}).
 The relation between $dr$
 and $dz$ is
\begin{equation}
dr = {\frac {c{\it dz}}{ \left( z+1 \right) H_{{0}}}}
\quad .
\end{equation}
The joint distribution in {\it z}
and      {\it f}  for the number of galaxies is
\begin{equation}
\frac{dN}{d\Omega dz df} =
\frac{1}{4\pi}\int_0^{\infty} 4 \pi r^2 dr \Phi (\frac{L}{L^*})
\delta\bigl(z-(e^{(H_0 \, r/c )} -1)\bigr)
\delta\bigl(f-\frac{L}{4 \pi r^2}    \bigr)
\quad ,
\label{nfunctionzlognornl}
\end{equation}
where $\delta$ is the Dirac delta function.

\begin{figure}
\begin{center}
\includegraphics[width=6cm]{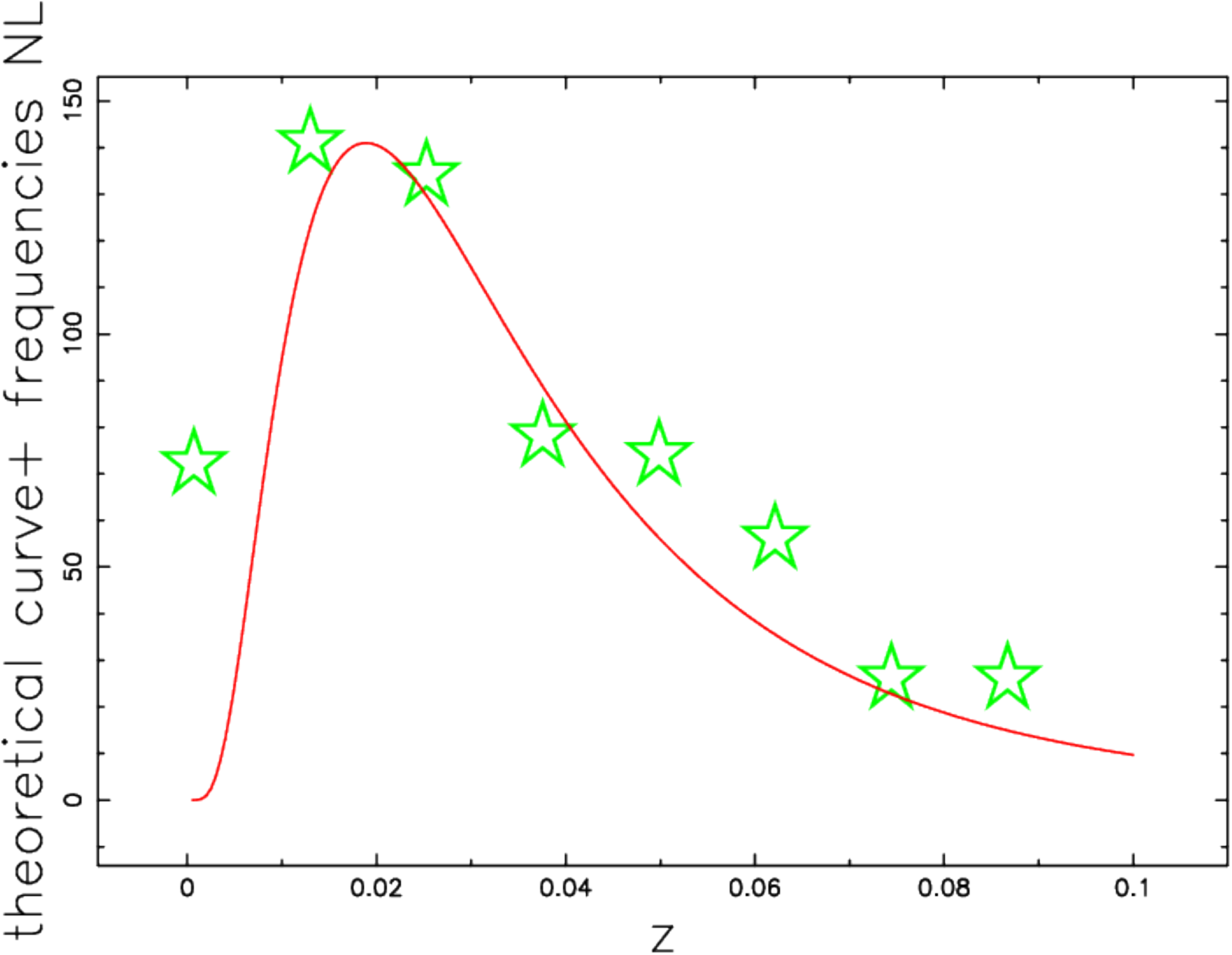}
\end {center}
\caption
{
Frequencies   of GRBs   at  a given  flux
as  a function  of the redshift, parameters as in Figure
\ref{max_pseudoeu}.
The full red line is the theoretical curve
generated by
$\frac{dN}{d\Omega dz df}(z)$
as given by the application of the lognormal LF
which  is eqn~(\ref{nfunctionzlognormnl}) in
the plasma  cosmology with
parameters as in Table \ref{coeflognormalsenza},
$\chi^2 =  6193$.
}
          \label{max_plasma}%
    \end{figure}

The evaluations of the integral  over luminosities and distances
gives
\begin{equation}
\frac{dN}{d\Omega dz df}=
\frac{
\left( \ln  \left( z+1 \right)  \right) ^{2}{c}^{3}\sqrt {2}{{\rm e}^
{-\frac{1}{2}\,{\frac {1}{{\sigma}^{2}} \left( \ln  \left( {\frac { \left(
\ln  \left( z+1 \right)  \right) ^{2}}{{z_{{{\it crit}}}}^{2}}}
 \right)  \right) ^{2}}}}{\it \Phi^*}
}
{
2\,\sqrt {\pi}{H_{{0}}}^{3}f\sigma\, \left( z+1 \right)
}
\quad  .
\label{nfunctionzlognormnl}
\end{equation}
The above formula has a maximum  at  $z=z_{pos-max}$, 
where
\begin{equation}
 z_{pos-max} =
{{\rm e}^{4\,{\frac {{\rm W} \left(\frac{1}{4}\,{\sigma}^{2}z_{{{\it crit}}}{
{\rm e}^{\frac{1}{2}\,{\sigma}^{2}}}\right)}{{\sigma}^{2}}}}}-1
\quad ,
\label{zmaxplasma}
\end{equation}
where $W(x)$ is the    Lambert $W$ function, see \cite{NIST2010}.
The above maximum can be re-expressed   as
\begin{equation}
 z_{pos-max} =
 {{\rm e}^{4\,{\frac {1}{{\sigma}^{2}}{\rm W} \left(\frac{1}{8}\,{\frac {{
\sigma}^{2}\sqrt {{\it L^*}}H_{{0}}{{\rm e}^{\frac{1}{2}\,{\sigma}^{2}}}}{
\sqrt {\pi}\sqrt {f}c}}\right)}}}-1
  \quad .
 \end{equation}
Figure \ref{max_plasma}
reports the
observed and theoretical number of GRBs
with a given flux
as a function of the  redshift.
In  the case of the plasma cosmology,
the theoretical maximum
as  given by eqn~(\ref{zmaxplasma})
is  at  $z=0.0188$, 
with the parameters as in Table \ref{coeflognormal},
against  the observed $z=0.019$.
The  theoretical averaged redshift of GRBs
is
$\langle z \rangle= 0.041$
 against the observed
$\langle z \rangle= 0.0385$.

\section{The truncated lognormal distribution}

This section derives the normalization and the mean
for a truncated lognormal PDF.
This truncated PDF fits  the high redshift behaviour
of the LF for GRBs.

\subsection{Basic equations}
\label{sectruncatedlognormal}

Let $X$ be a random variable taking
values $x$ in the interval
$[x_l, x_u ]$;
the truncated lognormal  (TL) PDF  is
\begin {equation}
TL(x;m,\sigma,x_l,x_u) =
\frac
{
\sqrt {2}{{\rm e}^{-\frac{1}{2}\,{\frac {1}{{\sigma}^{2}} \left( \ln  \left( {
\frac {x}{m}} \right)  \right) ^{2}}}}
}
{
\sqrt {\pi}\sigma\, \left( -{\rm erf} \left(\frac{1}{2}\,{\frac {\sqrt {2}}{
\sigma}\ln  \left( {\frac {x_{{l}}}{m}} \right) }\right)+{\rm erf}
\left(\frac{1}{2}\,{\frac {\sqrt {2}}{\sigma}\ln  \left( {\frac {x_{{u}}}{m}}
 \right) }\right) \right) x
}
\label{lognormaltruncated}
\quad .
\end {equation}
Its expected value   is
\begin{equation}
E(m,\sigma,x_l,x_u)
=
\frac
{
{{\rm e}^{\frac{1}{2}\,{\sigma}^{2}}}m \left( {\rm erf} \left(\frac{1}{2}\,{\frac {
\sqrt {2} \left( {\sigma}^{2}+\ln  \left( m \right) -\ln  \left( x_{{l
}} \right)  \right) }{\sigma}}\right)-{\rm erf} \left(\frac{1}{2}\,{\frac {
\sqrt {2} \left( {\sigma}^{2}+\ln  \left( m \right) -\ln  \left( x_{{u
}} \right)  \right) }{\sigma}}\right) \right)
}
{
{\rm erf} \left(\frac{1}{2}\,{\frac {\sqrt {2} \left( -\ln  \left( x_{{l}}
 \right) +\ln  \left( m \right)  \right) }{\sigma}}\right)-{\rm erf}
\left(\frac{1}{2}\,{\frac {\sqrt {2} \left( -\ln  \left( x_{{u}} \right) +\ln
 \left( m \right)  \right) }{\sigma}}\right)
}
\quad .
\label{meanlognormaltruncated}
\end{equation}
The distribution
function    is
\begin{equation}
DF(x;m,\sigma,x_l,x_u)=
\frac
{
-{\rm erf} \left(\frac{1}{2}\,{\frac {\sqrt {2}}{\sigma}\ln  \left( {\frac {x
}{m}} \right) }\right)+{\rm erf} \left(\frac{1}{2}\,{\frac {\sqrt {2}}{\sigma}
\ln  \left( {\frac {x_{{l}}}{m}} \right) }\right)
}
{
{\rm erf} \left(\frac{1}{2}\,{\frac {\sqrt {2}}{\sigma}\ln  \left( {\frac {x_{
{l}}}{m}} \right) }\right)-{\rm erf} \left(\frac{1}{2}\,{\frac {\sqrt {2}}{
\sigma}\ln  \left( {\frac {x_{{u}}}{m}} \right) }\right)
}
\quad .
\label{dftruncatedlognormal}
\end{equation}
The four parameters which characterize  the truncated
lognormal distribution 
can be found with the 
maximum likelihood estimators  (MLE) and by the evaluation 
of the minimum and maximum elements
of the sample.
The  LF  for   GRB as given
by the truncated lognormal,
$\Phi_T (L;L^*,\sigma,L_l,L_u)$,
is  therefore
\begin{equation}
\Phi_T (L;L^*,\sigma,L_l,L_u) =\Phi^*
TL(L;L^*,\sigma,L_l,L_u)
\frac{number}{Mpc^3 \,yr}
\quad ,
\label{pdflognormaltruncgrb}
\end{equation}
where $L^*$ is the scale,
$L_l$ the lower bound in luminosity,
$L_u$ the upper bound in luminosity
and $\Phi^* $ is given by eqn~(\ref{fistar}).

\subsection{Applications at high $z$}

\begin{figure}
\begin{center}
\includegraphics[width=10cm]{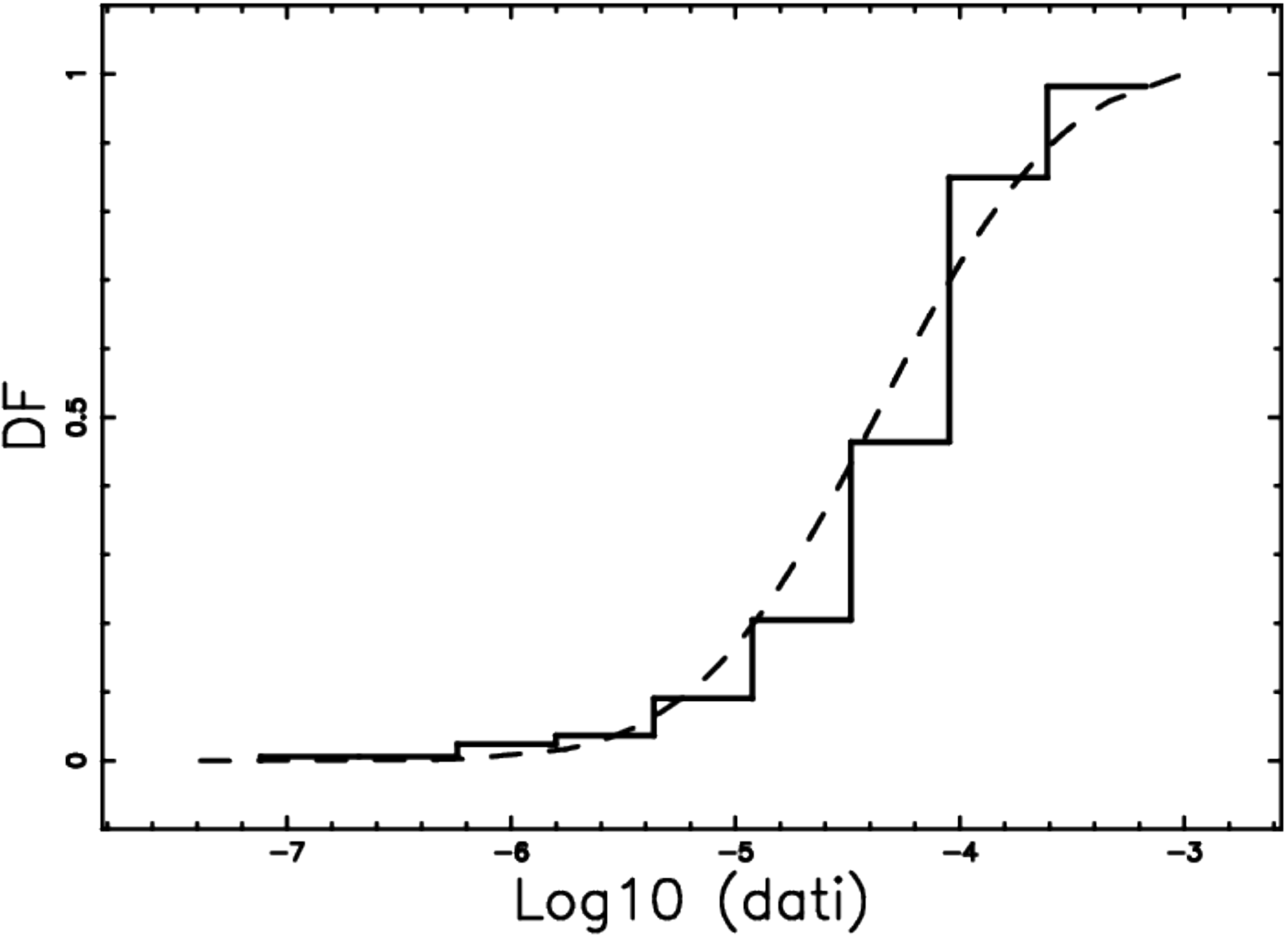}
\end {center}
\caption
{
Observed DF (step-diagram)
for  GRB luminosity and  superposition
of the truncated lognormal DF  (line),
case of the $\Lambda$CDM cosmology with  parameters as
in Table \ref{coeflognormaltronc}.
}
\label{lognorm_lcdm_tronc}
    \end{figure}

\begin{figure}
\begin{center}
\includegraphics[width=10cm]{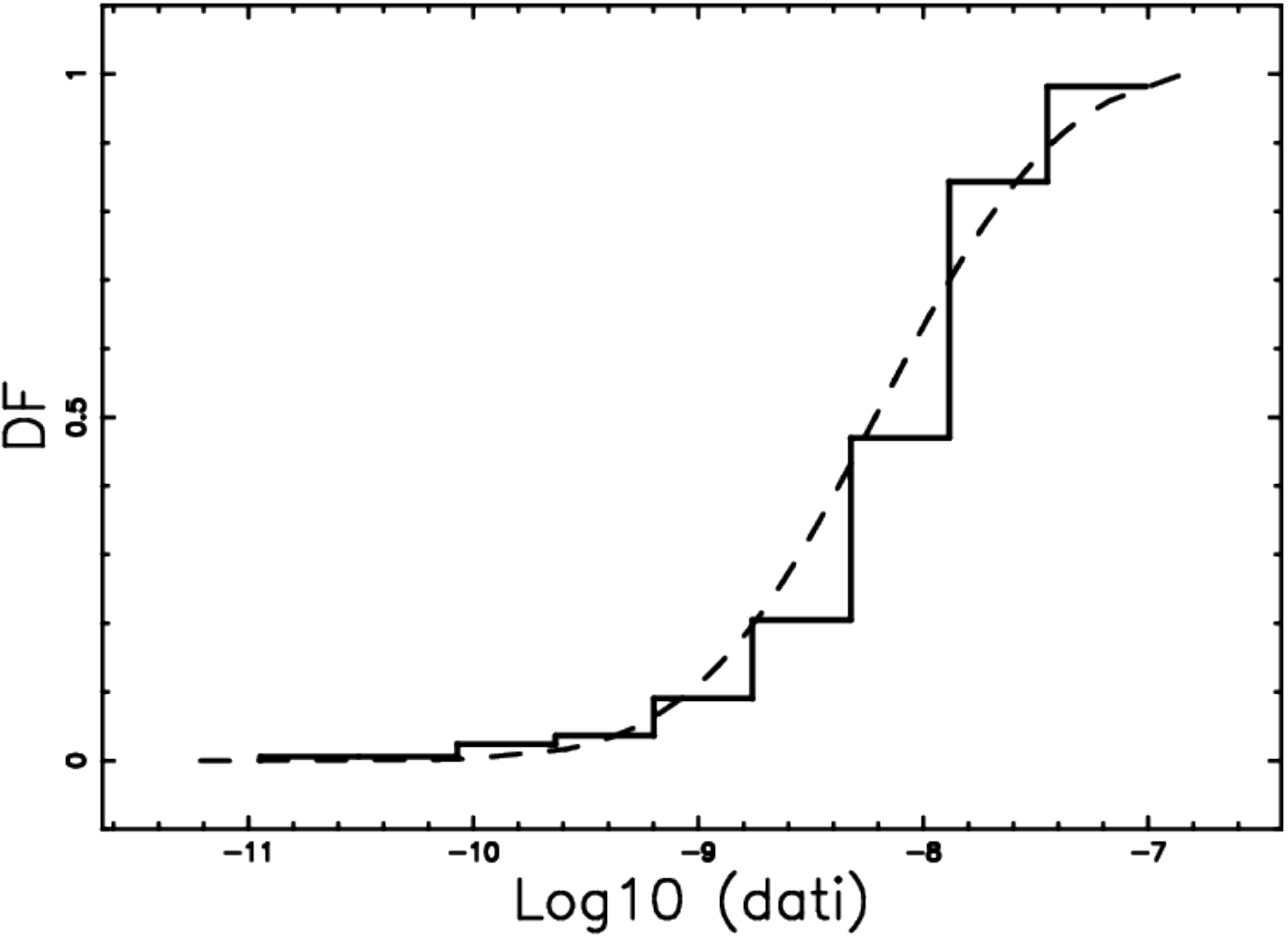}
\end {center}
\caption
{
Observed DF (step-diagram)
for  GRB luminosity and  superposition
of the truncated lognormal DF  (line),
the case of the Plasma  cosmology without $k(z)$
correction with  parameters as
in Table \ref{coeflognormaltronc}.
}
\label{lognormal_df_tronc_plasmasenza}
    \end{figure}

The LF for GRBs as modeled by a truncated lognormal DF
is reported  in Figure \ref{lognorm_lcdm_tronc}
in the case
of the $\Lambda$CDM cosmology
and  in
Figure \ref{lognormal_df_tronc_plasmasenza}
in the  case of the plasma cosmology without a $k(z)$ correction;
the data is as in Table \ref{coeflognormaltronc}.
\begin{table}[ht!]
\caption
{
The 5 parameters of  the LF as modeled by the
truncated lognormal distribution
for  $z$ in $[0,0.02]$
and  the two parameters of the K--S test
$D$ and  $P_{KS}$.
We analysed
the case of
the $\Lambda$CDM cosmology
where
the luminosity is  given by
eqn~(\ref{loglumlcdm}) in the second column
and the case of the plasma cosmology,
the case of the  X-band (14--195 keV) without $k(z)$ correction,
where the  luminosity is  given by
eqn~(\ref{loglumplasmasenza}), third column.
}
\label{coeflognormaltronc}
\begin{center}
\begin{tabular}{|c|c|c|c|}
\hline
Parameter  &  $\Lambda$CDM cosmology & Plasma~cosmology \\
\hline
$\frac{L_l}{10^{51} erg\,s^{-1}}$  & $4.11\,10^{-8}$   & $6.11\,10^{-12}$                    \\
$\frac{L_u}{10^{51} erg\,s^{-1}}$  & $9.8 \,10^{-4}$    & $1.42\,10^{-7}$                  \\
$\frac{L^*}{10^{51} erg\,s^{-1}}$    & 4.05\,$10^{-5}$ & $5.9\,10^{-9}$  \\
$\sigma$ & 1.42   &   1.42                        \\
$\frac
{\Phi^*}
{Mpc^{-3} yr^{-1}}
$ & $1.02\,10^{-5}$ & $1.01\,10^{-5}$               \\
$D$             & 0.084 & 0.084       \\
$ P_{KS}$       & 0.177 & 0.18   \\
\hline
\end{tabular}
\end{center}
\end{table}

In order to model evolutionary effects,
a variable upper bound in luminosity, $L_u$,
has been introduced
\begin{equation}
L_u = 1.25 (1+z)^2 \,10^{51} \quad \frac{erg}{s}
\label{luobservedmax}
\quad ,
\end{equation}
see eqn~(7) in \cite{Tan2013};
conversely
the lower bound, $L_l$  was already fixed
by eqn~(\ref{loglumlcdm}).
A second evolutionary correction  is
\begin{equation}
\sigma =\sigma_0 (1+z)^2
\quad ,
\label{sigmacorrection}
\end{equation}
where  $\sigma_0$ is the evaluation  of $\sigma$
at  $ z \approx 0$, see Table \ref{coeflognormaltronc}.

\begin{figure}
\begin{center}
\includegraphics[width=10cm]{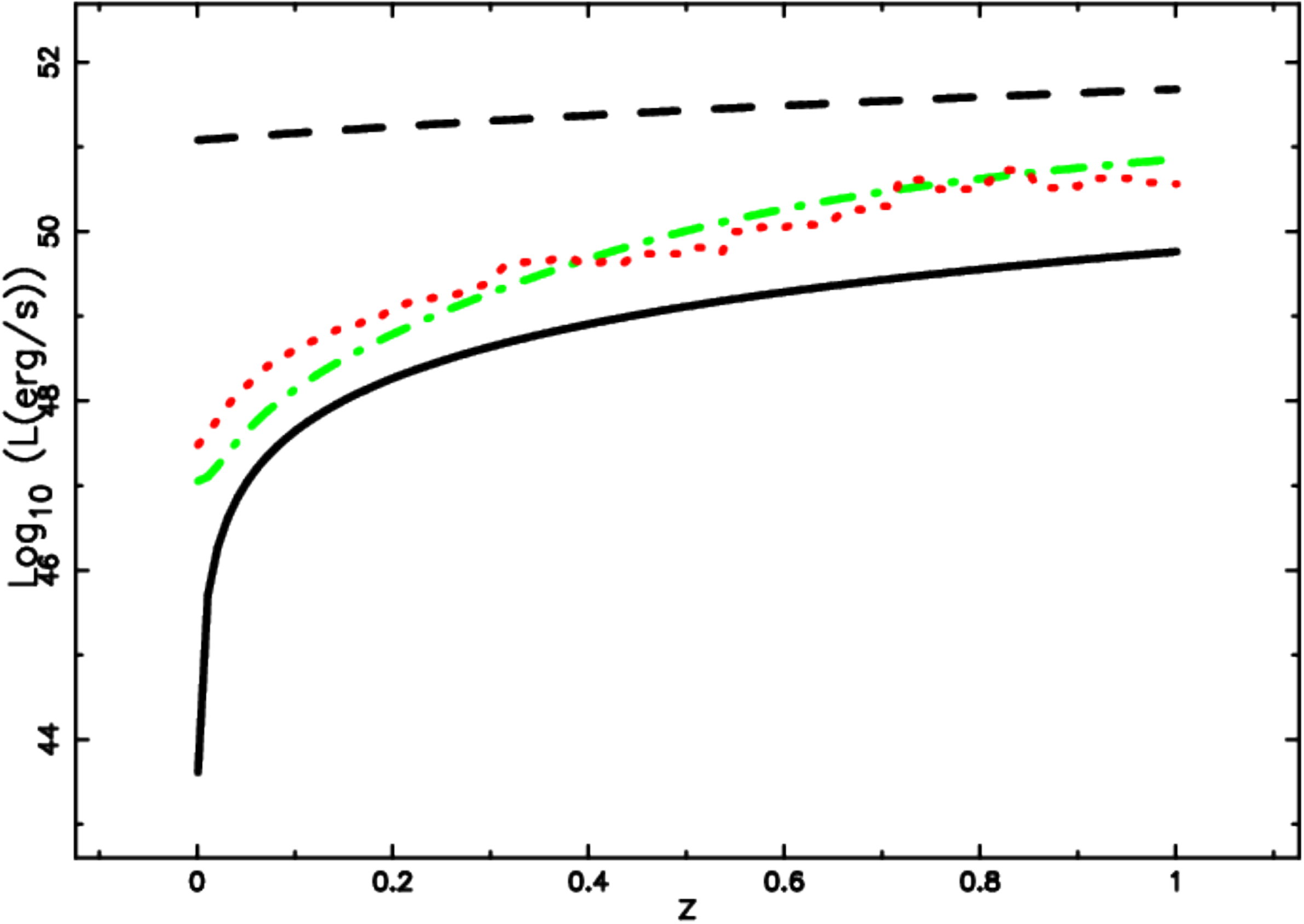}
\end{center}
\caption
{
Average observed luminosity in the $\Lambda$CDM cosmology
versus redshift for 784 GRB (red points),
theoretical average luminosity for truncated lognormal LF
as given by eqn~(\ref{meanlognormaltruncated})
(dot-dash-dot green line),
theoretical curve for the lowest luminosity at  a given
redshift,
see eqn~(\ref{loglumplasmasenza}) (full black line) and
the empirical curve
for the highest luminosity at  a given
redshift (dashed black line),
see eqn~(\ref{luobservedmaxplasma}).
}
\label{lognorm_lcdm_tronc_xm}
\end{figure}

Figure \ref{lognorm_lcdm_tronc_xm}
reports a comparison between
the theoretical average luminosity
and the observed average luminosity for
the $\Lambda$CDM cosmology.

\begin{figure}
\begin{center}
\includegraphics[width=10cm]{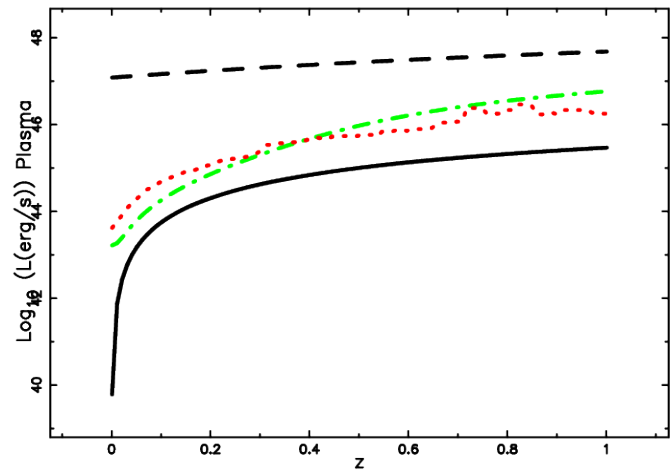}
\end{center}
\caption
{
Average observed luminosity in the Plasma  cosmology without $k(z)$
correction versus redshift for 784 GRB (red points),
theoretical average luminosity for truncated lognormal LF
as given by eqn~(\ref{meanlognormaltruncated})
(dot-dash-dot green line),
theoretical curve for the lowest luminosity at  a given
redshift,
see eqn~(\ref{loglumlcdm}) (full black line) and
the empirical curve
for the highest luminosity at  a given
redshift (dashed black line),
see eqn~(\ref{luobservedmaxplasma}).
}
\label{lognorm_plasma_tronc_xm}
\end{figure}

In the case of the plasma cosmology,
the  variable upper bound in luminosity, $L_u$,
is
\begin{equation}
L_u = 1.25 (1+z)^2 \,10^{47} \quad \frac{erg}{s}
\label{luobservedmaxplasma}
\quad ,
\end{equation}
and
Figure \ref{lognorm_plasma_tronc_xm}
reports a comparison between
the theoretical average luminosity
and
the observed average luminosity for the plasma  cosmology.

\section{Conclusions}

{\bf Luminosity}

The evaluation of the luminosity
is connected  with the evaluation of the luminosity distance,
which is different for every adopted cosmology:
the $\Lambda$CDM and plasma cosmologies cover the range in $z$
$[0-4]$ and  the pseudo-Euclidean cosmology
covers the limited range in $z$, $[0-0.15]$.

An application of a  correction for the luminosity over all
the $\gamma$ range which is $[1\,kev-10^4\,kev]$
allows speaking  of the extended luminosity of a GRB;
in the case of  $\Lambda$CDM, see eqn~(\ref{loglumlcdm}),
which depends on the three observable parameters
$fluxfwm2$, $z$  and $\gamma$.
An analytical formula for the luminosity in $\Lambda$CDM without corrections
is given as a function of the two observable parameters
$fluxfwm2$ and  $z$, see eqn~(\ref{lumlcdmmono}),
which can be tested on
the SWIFT-BAT catalog
of \cite{Baumgartner2013}.

{\bf Lognormal luminosity function}

We analysed the widely used   lognormal
PDF as a LF for GRBs, see Section \ref{seclognormal}.
We derived an expression for the maximum
in the number of GRBs for a given  flux,
which is  eqn~(\ref{zmaxpseudo})  in the linear
case (pseudo-Euclidean universe), see
also Figure \ref{max_pseudoeu},
and eqn~(\ref{zmaxplasma})  in the non-linear
case (plasma cosmology), see
also Figure \ref{max_plasma}.

{\bf Four broken power law luminosity function}

The four broken power law PDF gives the best  statistical results
for the LF of GRBs, see Table \ref{fourbroken}.
The  weak point  of this LF  is in the number of parameters,
which is 
9, against the 4 of the
truncated lognormal LF or 
2 of the  lognormal LF. 

{\bf Maximum in flux}

The maximum in the joint distribution in redshift and energy flux density 
is here modeled in the case of a pseudo-Euclidean universe
adopting a standard technique  originally developed for galaxies,
see formula (5.132) in \cite{Peebles1993}
and our formula (\ref{dnsudz_pseudo}).
In the case of the plasma cosmology, the maximum has been found by analogy, see 
our formula (\ref{dnsudz_pseudo}).
In the case of the $\Lambda$CDM cosmology, the redshift as a function of
the luminosity has a complex behaviour, see 
formula (66) in \cite{Zaninetti2015a}, and the
analysis has been postponed to future research.
The above complexity has been considered in an example
of a simpler plasma cosmology rather than  
in the $\Lambda$CDM cosmology.

{\bf Evolutionary effects}

The LF for GRBs at high $z$ is well modeled
by a truncated lognormal PDF,
see Section \ref{sectruncatedlognormal}.
The lower bound for the  luminosity is fixed
by  the decrease in the range of observable
luminosities and the
higher bound by a standard assumption,
see eqn~(\ref{luobservedmax}).
A further refinement of the truncated lognormal
model for the GRBs at high $z$ is obtained by
introducing a cosmological correction
for $\sigma$, see eqn~(\ref{sigmacorrection}),
see Figure \ref{lognorm_lcdm_tronc}
for the case of the $\Lambda$CDM cosmology
and  Figure \ref{lognorm_plasma_tronc_xm}
for the case of the plasma cosmology.
In other words, the $\Lambda$CDM cosmology
and the plasma cosmology are indistinguishable
in the range of redshifts here analysed, 
$0\,\leq\,z\leq 4$. 


\end{document}